\documentstyle[multicol,aps,prb,epsf]{revtex}
\begin{document}

\draft

\title{Electronic structure of periodic minimal surfaces\\
--- `topological band structure'}
\author{\noindent H. Aoki, M. Koshino, D. Takeda\cite{Takeda}, H. Morise}
\address{Department of Physics, University of Tokyo, Hongo, Tokyo
113-0033, Japan}
\author{K. Kuroki}
\address{Department of Applied Physics and
Chemistry, University of Electro-Communications, Chofu, 
Tokyo 182-8585, Japan}

\date{\today}

\maketitle

\begin{abstract}
  Electronic band structure for electrons bound on periodic 
  minimal surfaces is differential-geometrically formulated and 
  numerically calculated.  
  We focus on minimal surfaces 
  because they are not only mathematically elegant 
  (with the surface characterized completely 
  in terms of ``navels") but represent the topology of real
  systems such as zeolites and negative-curvature fullerene.  
  The band structure turns out to be primarily determined by the topology of
  the surface, i.e., how the wavefunction interferes on a
  multiply-connected surface, so that the bands are 
  little affected by the way in which we 
  confine the electrons on the surface (thin-slab limit or 
  zero thickness from the outset).  
  Another curiosity is that different minimal surfaces 
  connected by the Bonnet transformation 
  (such as Schwarz's P- and D-surfaces) possess 
  one-to-one correspondence in their band energies 
  at Brillouin zone boundaries.  
   
\end{abstract}



\begin{multicols}{2}
\narrowtext

\section{Introduction}
There is a long history for the fascination in particles bound on curved
surfaces, which dates back to the early days of quantum  
mechanics\cite{dewitt}.  There are two complimentary points of interests:  
one is how the particle motions are affected by the local 
curvature of the surface.  The other is how the global 
topology (i.e., how the surface is wound) affects 
quantum mechanical wave functions.  
The latter problem becomes especially interesting if we 
consider a {\it periodic} surface embedded in the three-dimensional 
space.  

Geometrically,  Schwarz, back in the 19th century, showed
that we can make curved surfaces extend over the 
entire three-dimensional space
by connecting hyperbolic (i.e., everywhere negatively curved)
patches. 
Specifically, Schwarz has constructed  
{\it periodic minimal surfaces}, 
where minimal means that the negatively-curved 
surface has a minimized area with the mean
curvature ($\frac{1}{2}(\kappa_1+\kappa_2)$ with $\kappa_1,\kappa_2$
being the principal curvatures) vanishing everywhere on the
surface.\cite{DiffG,mackayperi}

There are reasons from both mathematics and condensed-matter physics
why periodic surfaces are intriguing.  
First of all, periodic surfaces are of general interest
in the condensed-matter physics.  
(i) In general, ``crystals'' (periodic structures) composed of
surfaces are conceptually interesting as a class of 
periodic system on which electrons move.
Mackay\cite{Mackay1,Mackay2,enume} has classified them
group-theoretically
(just as the ordinary crystals composed of atoms are classified
with the space group), which he named `flexi-crystallography'.

(ii) In terms of materials science, periodic minimal surfaces 
represent the topology
of real condensed-matter systems.  These
include not only conventional materials such as zeolites or
a silica polymorph called melanophlogite\cite{melano,PMM}, or
isostructural silicon clathrates\cite{moriguchi},
but recent advances in fabrication of exotic materials such as
fullerenes or nanotubes have inspired further possibilities
such as negative-curvature fullerene
(or C$_{60}$ zeolite)\cite{MacTerr,Tersoff,OKeef,Lenosky,Fujita}
whose fabrication has been attempted with a zeolite as a
template\cite{kyotani}.
Their structures can be modeled as curved surfaces
if we smear out atoms into a surface in the effective-mass sense, 
and it is a fundamental
question to consider how a mobile (e.g., $\pi$)
electron behaves on such surfaces

Second, there are mathematical interests and simplifications 
when a periodic surface is minimal:
we can exploit the Weierstrass representation, which enables
us to specify the surface in a surprisingly simple manner
in terms of ``navels".  The representation also
simplifies Schr\"{o}dinger's equation as we shall show in 
the present paper.  

There are further mathematical fascinations specific to surfaces.  
One virtue of the structure constructed from surfaces is we can deform it.  
One can in fact deform one minimal surface into 
another with a differential geometrical transformation 
called the Bonnet transformation.  We can then raise a 
question of how the band structure for one surface 
could be related with that for the transformed one.  
Another interest is that some periodic minimal surfaces,
such as Schwarz's P-surface\cite{Schwarz}, have a high symmetry
(`interior-exterior' symmetry) that divides the space into two
equivalent parts, which should be reflected in the electronic band
structure.  

So in the present paper we address ourselves a question: 
how the electronic band structures should 
look like for periodic minimal surfaces. 
To start with, however, we have to envisage there are in general two
ways (Fig.\ref{fig:Schematic}) 
to make electrons confined to a surface: (a) One is to
consider electrons bound to a thin, curved slab of thickness $d$, where
the limit $d\rightarrow 0$ is taken\cite{Nagaoka}.  (b) The other is 
to consider the surface with the degree of freedom normal to the surface
ignored from the outset, i.e., a two-dimensional sheet is rolled into
the curved surface\cite{Ogawa}. 
Either way it has been shown that an effective potential arises from the
curvature of the surface, but that the potential is different between 
the two cases.   
Namely, the thin slab case (a)\cite{Nagaoka} has a potential 
\[
-(\hbar^2/8m)(\kappa_1 - \kappa_2)^2, 
\]
while the case (b)\cite{Ogawa} has 
\[
+(\hbar^2/8m)(\kappa_1 + \kappa_2)^2.
\]
The origin of the discrepancy
was subsequently revealed by Nagaoka and coworkers:\cite{Takagi} when
the degree of freedom normal to the surface is ignored, Dirac's
prescription for constrained systems can be applied, but there is a
room for ambiguity in the order of operators.  If we adopt the
conservation constraint, the resultant equation reduces to that in the
$d\rightarrow 0$ approach.  
When the surface is minimal
($\kappa_1+\kappa_2=0$), the curvature potential is 
nonzero in general (since $\kappa_1-\kappa_2=2\kappa_1\neq 0$) 
in case (a), while the curvature
potential vanishes identically in (b).  

For condensed-matter systems such as atoms
arrayed along a curved surface, we should take the
$d\rightarrow 0$ approach.  Still, the difference 
in the band structure between the two cases is 
curious.  Namely, although we have a periodic system 
in either case, the periodicity imposed in case (a) is
the periodicity in the strong potential 
that confines the electron into a thin slab (Fig.1(a)), while 
in the case (b) the electron moves freely along the surface, 
where the only constraint is that an electron has to move
in a space having a nontrivial topology.  The topology 
can have a profound effect on the electron's wave function, since, 
if we regard the periodic surface as a network of pipes
(a cubic network for P-surface, diamond for D-surface, etc),
the wavefunction interferes with itself along various paths
wound around the ``necks".  
Thus the periodicity felt by an electron amounts to the strong confining 
potential in case (a), while the periodicity only
enters as a way in which the wave function interferes in case (b),
and it is an intriguing question whether or not their band structures 
are similar.

The purpose of this paper is, (i) to explicitly write down
Schr\"{o}dinger's equation for electrons on periodic minimal surfaces
by exploiting Weierstrass's representation in order to obtain the
electronic band structure; (ii) to calculate and compare the 
band structures in cases (a) and (b).
Unexpectedly the band structures turns out to be similar between cases
(a) and (b), i.e., the bands are primarily determined from the 
topological way in which the wave function interferes with itself.  
The energy scale of the band structure (band splitting,
such as a split of the `d' band into E$_g$ and T$_{2g}$, and band
widths) is also universally $\sim \hbar^2/2mL^2$ with $L$ being the linear
dimension of the unit cell of the periodic surface.  
(iii) We go 
further to `Martensitic-deform' a surface 
to another connected by the Bonnet 
transformation.  We shall show that 
there exists a curious one-to-one correspondence in their 
band structures, which illustrates another curious feature 
in the topological band structure.

\section{Weierstrass representation of minimal surfaces}
We start with a mathematical prerequisite for representing minimal
surfaces.   A two-dimensional surface ${\bf r}(q^1,q^2)$
embedded in a three-dimensional space
can be expressed in terms of two dimensional coordinates $q^1,q^2$, where
$(q^1,q^2)\equiv (u,v)$ are called isothermal when the metric
tensor $g_{ij}$ is diagonal with
\[
{\rm \, d} {\bf r}\cdot{\rm \, d}{\bf r} = g_{11} ({\rm \, d} u {\rm \,
d} u + {\rm \, d} v {\rm \, d}v).
\]
What Weierstrass and Enneper have found is that
a necessary and sufficient condition for ${\bf r}(u, v)$
representing a minimal surface with isothermal $(u,v) \in S$
($S$: a simply connected region) is that
there exist $F, G$, functions of $w \equiv u + i v$,
with which ${\bf r}(u,v) = (x(u,v),y(u,v),z(u,v))$ is expressed as
\begin{eqnarray}
  {\bf r}(u, v) = {\rm Re} \left(
    \int^w_{w_0} F (1 - G^2) {\rm \, d}w,  \right. \nonumber \\
  \left.
    \int^w_{w_0} i F (1 + G^2) {\rm \, d}w,
    \int^w_{w_0} 2F G {\rm \, d}w
  \right),
  \label{eqn:Weiermap}
\end{eqnarray}
where $w_0$ is a constant,
and $FG^2$ is assumed to be regular
(i.e., $m$-th poles of $G$ assumed to coincide with $2m$-th
zeros of $F$).\cite{Mackay1,Mackay2,Terrones}
If there are singularities that violate this condition,
we can exclude these points by incising cut(s) to make
$S$ a Riemann surface.
Thus there is a one-to-one correspondence between a minimal
surface and the functional form of $F,G$.

Now, Schr\"{o}dinger's equation for a curved
surface, expressed with two dimensional coordinates $(q^1,q^2)$ 
and metric tensor $g_{ij}$, is written\cite{} as
    \begin{eqnarray}
      \left[  - {\hbar^2 \over 2m}
        {1 \over \sqrt{g}}
        {\partial \over \partial q^i} \sqrt{g}\ g^{ij}
        {\partial \over \partial q^j}
        - {\hbar^2 \over 8m}\left( \kappa_1 - \kappa_2 \right)^2
      \right] \psi(q^1, q^2) \nonumber = \\ E\ \psi(q^1, q^2),
      \label{eqn:SchResult2}
    \end{eqnarray}
where summations over repeated indices are assumed.  
This equation is for model (a), while we can replace 
the second term in the bracket (potential term) by 
$+{\hbar^2 \over 8m}\left( \kappa_1 + \kappa_2 \right)^2$
for model (b). 

In the Weierstrass-Enneper representation, every quantity in the 
Schr\"{o}dinger's equation can be expressed in terms of $F$ and $G$, since 
the Laplacian, the first term in the angular
brackets in eq.\ref{eqn:SchResult2}, 
reduces to $(\partial^2/\partial q_i^2)/\sqrt{g}$ in the
isothermal coordinates, where
\[
g\equiv {\rm det}\{g_{ij}\}=
\left[\frac{1}{2}|F|(|G|^2+1)\right]^4 ,
\]
while we can plug in $\kappa_1 = -\kappa_2 =
4|G^\prime|/|F|(|G|^2 + 1)^2$ for the curvature term.
Schr\"{o}dinger's equation for periodic minimal surfaces 
then reduces to
\begin{equation}
 -\frac{4}{|F|^2 (|G|^2 + 1)^2 }
      \left[
              {\partial^2 \over \partial u^2}
	      + {\partial^2 \over \partial v^2}
	      + \frac{4|G^\prime|^2}{(|G|^2 + 1)^2}
      \right] \psi = \varepsilon \ \psi,
\end{equation}
where $\varepsilon \equiv E/(\hbar^2/2m)$.
As evident from the Weierstrass representation (\ref{eqn:Weiermap}),
$F$ has the dimension of length and $G$ is dimensionless.
Hence the energies in minimal surfaces always scale as 
$E/(\hbar^2/2mL^2)$, where 
$L \sim F \sim$ linear dimension of the unit cell 
(a precise expression given in eqn.\ref{eqn:WEP} below).

We can have a more transparent form when $G(w)=w$ (as 
is often the case with periodic minimal surfaces, including P-surface).  
In this case we can 
exploit the stereographic map (Gauss map) from the infinite 
complex plane $(u,v)$ to a unit sphere $(\theta, \phi)$,
\[
w=u+iv=\cot\left(\frac{\theta}{2}\right) {\rm e}^{i\phi}.
\]
After a bit of algebra, we finally arrive at the differential
equation for $(\theta, \phi)$,
    \begin{equation}
      -{ (1 - \cos\theta)^4 \over |F|^2 }
      \left(
              {\partial^2 \over \partial\theta^2}
              + \cot\theta{\partial \over \partial\theta}
               + {1 \over \sin^2 {\theta}}{\partial^2 \over
\partial\phi^2}
               +  1
      \right) \psi = \varepsilon \ \psi .
      \label{eqn:SchOnP}
    \end{equation}

Curiously, a common coefficient $(1 - \cos\theta)^4/|F|^2$ factors out
for the Laplacian (the first three terms in the large parentheses in the
above equation) and the curvature potential (+1), which 
is made manifest due
to the Weierstrass representation.  Hence the kinetic and potential
energies vary in a correlated manner as we go from one minimal surface
to another by changing $F$.  
However, this does not imply that the curvature potential
always exerts as large an effect as the kinetic energy does, 
since the expectation values of the kinetic
and potential energies depend on the amplitude of the wave function, and
a quantitative study is required.

\section{Results for Schwarz's P-surface}
So we first take Schwarz's P-surface as a typical triply-periodic
(periodic in $x,y,z$) minimal surface, or a `simple cubic' minimal
surface (belonging to a space group Im3m in the language of the
flexi-crystallography). Some authors\cite{Terrones,Cvijovic,Klinowski}
gave the Weierstrass-Enneper representation for the P-surface as 
\begin{equation}
F(w) = iL/\sqrt{1 - 14w^4 + w^8},
\label{eqn:WEP}
\end{equation}
where $L \sim$ linear dimension of the unit cell 
(to be precise the unit cell size is $2.157L$, 
which is given as an elliptic integral).

A unit cell of the P-surface comprises eight identical 
patches, as shown in Fig.\ref{fig:SurfaceUnit}.
With the stereographic mapping discussed above, a unit cell is 
mapped onto two spheres, connected
into a Riemann surface via four cuts, which have to be introduced to
make $FG^2$ non-singular since $F$ has poles.  

Differential geometrically $F$ is in general 
specified solely by such poles in a form $F \propto 1/\prod_i(w-w_i)^\eta$ where
$\eta$ determines the topology of the surface.
In fact the poles correspond, in the language of differential geometry, 
to the
{\it navels} (umbilical points), 
which are defined as the points where
every cross section is inflected, with 
the two principal curvatures becoming degenerate, $\kappa_1=\kappa_2$
($=0$ for a minimal surface).
So a periodic minimal surface is completely characterized by
the navels that appear periodically.
The curvature potential $\propto -(\kappa_1-\kappa_2)^2$
in case (a) varies on the surface,
which may be called a `crystal field' in flexi-crystals.
Navels then specify the positions where the curvature potential becomes
maxima ($=0$), while
the minima occurs at the maxima of the absolute value of $\kappa_1(=-\kappa_2$ 
for a minimal surface).  
In the P-surface the navels (potential maxima) occur at eight
`Affensattel' (monkey's-saddle) points in a unit cell\cite{Hilbert}, while 
the potential minima 
occur at four points around each nape of the neck 
as depicted in Fig.\ref{fig:SurfaceUnit}(a).\cite{foam}

\section{Results for the band structure}
{\it Band structure for case (a)} \hspace*{0.5cm} 
In Schr\"{o}dinger's equation for
periodic minimal surfaces, eqn.(\ref{eqn:SchOnP}), the variables $\theta,
\phi$ cannot be separated, so that we have solved the equation
numerically by discretizing $\theta, \phi$ to diagonalize the
Hamiltonian matrix.  In discretizing the spherical coordinates, a
special care is taken around the navels, since the Jacobian $J$ of the
transformation to the Gauss sphere is singular there (Fig.\ref{fig:pointP}).  
The band
structure is obtained by connecting the adjacent unit cells with
appropriate phase factors.

We now come to the result for the band structure for the P-surface in
Fig.\ref{fig:BandP} (curves), 
and typical wavefunctions at $\Gamma$ and H in the bcc Brillouin 
zone ($\Gamma$ in the simple-cubic zone) 
in Fig.\ref{fig:WaveP}. The P-surface happens to divide the space into two
equivalent parts, since a body center enclosed by the 
surrounding unit cells has the same shape as the original unit, so that
we first note that the true symmetry is body-centered cubic rather than
simple cubic.  So the bands are displayed on the Brillouin zone for bcc.
In accord with the above argument, the energy scale (band width, splitting
etc) is $\sim \hbar^2/2mL^2$.  This is of the order of $1$ eV for 
$L\sim 10$ \AA, the unit cell size 
assumed for a hypothetical negative-curvature fullerene\cite{Townsend}.

The curvature, or the effective mass, of these bands are
either positive (electron-like) or negative (hole-like)
according to the nature of the wave function.
The mass cannot be estimated with a simple ${\bf k}\cdot {\bf p}$
perturbation, since the perturbation $\propto{\bf k}\cdot {\bf p}$
derives from the fact that $H_0\propto {\bf p}^2$ while
$H_0$ has no such simple form on a curved surface.
In other words,
the ${\bf k}\cdot {\bf p}$ formula has 
$\sum_j (\langle i|p_{\mu}|j \rangle \langle j|p_{\nu}|i
\rangle)/(E_i-E_j)$,
so we would have to calculate the matrix elements of ${\bf p}$
for wavefunctions that are finite only along the surface.

{\it Band structure for case (b)} \hspace*{0.5cm} 
Now we are in position to compare the (a) confinement case 
(curves in Fig.\ref{fig:BandP}) 
with the (b) rolled case (dotted lines in Fig.\ref{fig:BandP}).  
We can immediately see that the two band structures are
rather similar up to some offset $(2.34\hbar^2/2mL^2)$.
This is surprising, since there is no apriori reason why
they should be.
To be more precise quantitative features characterizing the
band, i.e., the effective mass and band widths,
are similar between the two cases.
So we conclude that the band structure does not essentially depend on
the way in which electrons are confined, at least for
`gently' curved surfaces such as minimal surfaces
where the there are no sharp edges that would give large
curvature potentials.

\section{Bonnet transformation} 

The next important question is: are the band structures for surfaces
connected by the Bonnet transformation related?  

{\it Bonnet transformation} \hspace*{0.5cm}
The deformation of the P-surface to other
periodic minimal surfaces can be implemented by the Bonnet 
transformation, which is conformal and 
is represented by an elliptic transformation.  
A beautiful asset of the Weierstrass representation for 
minimal surfaces 
is that the Bonnet transformation is simply represented by 
a phase factor, $F \rightarrow Fe^{i\beta}$, in 
eqn.(\ref{eqn:Weiermap}), where $\beta$ is called the Bonnet angle.  If
we apply this to the P-surface (cubic), the transformation
changes\cite{Terrones} it into the G-surface (gyroid with $\beta=0.211\pi$) and
the D-surface (diamond with $\beta=\pi/2$), which may be regarded as a
`Martensitic transformation' in the words of Ref.\cite{hyde}.  The
structure of the D-surface is depicted, along with the P-surface, in
Fig.\ref{fig:SurfaceUnit}.  Since the D-surface has a diamond symmetry, 
its unit cell contains two `cages'.  

We can first note that the Bonnet transformation preserves 
the metric tensor
and the Gaussian curvature. This implies that all the
surfaces connected by the Bonnet transformation obey the {\it identical}
Schr\"{o}dinger equation within a unit patch.  
Indeed, if we look at Schr\"{o}dinger's
eqn.(\ref{eqn:SchOnP}), $F$ only
enters as $|F|$, so that $F\rightarrow Fe^{i\beta}$ does not alter the
equation.  Although this is curious enough, this does not mean that the band
structures are identical, since, while the transformed surfaces share the
same genus ($=$ three for P and D), the way in which unit cells are
connected is different among them.

{\it Band structure of a Bonnet-transformed surface} \hspace*{0.5cm}

Figure \ref{fig:BandP_D} compares the band structures for the 
P- and D-surfaces.  
The two band structures are indeed different due
to the difference in three dimensional connection of the unit cells 
discussed above.  Curiously, 
however, we find that the values of the band energy at {\it special
points} (Brillouin zone corners, edges and face-centers) have identical
set of values between different surfaces.  
Namely, a close comparison of the two band structures reveals that 
the band energies exactly coincide, where the 
`law of correspondence' is\par
\begin{tabular}{ccc}
P-surface & & D-surface \\ \hline
$\Gamma$, H & $\Leftrightarrow$  & $\Gamma$, R \\
N           & $\Leftrightarrow$  & X, M        \\
\end{tabular}

This can be explained from the property of the 
Bonnet transformation that does not change Schr\"{o}dinger's 
equation.  For this purpose we have to look at the unit cells more 
closely.  
In Fig.\ref{fig:Diagram}(a), we show how unit patches are connected 
for the D- and P-surfaces.  We have indicated in the figure 
how the eight patches in a unit cell, numbered with 1 through 8, 
are connected to other patches in the adjacent cells by marking the edges 
with those numbers (i.e., if an edge is marked with, say, 7, the adjacent 
patch should be 7).   The wave function should be continued 
to the adjacent patch with a certain phase factor.  
We have indicated  the connection coefficients,
\[
\rho_i = \exp(i\phi_i),
\]
where $\phi_i$ is the Bloch phase along $i=x,y,z$.   
Hence this diagram fully characterizes 
how the Bloch wave functions are connected 
on the periodic minimal surfaces in terms of patches. 
We can then compare the 
coefficients for the P- and D-surfaces to
extract a correspondence at special points in $k$-space.

The diagrams introduced here have naturally different 
patch numbers assigned on them between the P- and D-surfaces, 
since the way in which the 
patches are connected (i.e., numbers attached to the edges) is 
different between them.  
However, we can make them identical, 
if we rearrange the connection numbers by 
noting the symmetry.  Since the parity 
inversion $\sigma_p$ with respect to the center of a unit cell preserves
the Hamiltonian and does not change the special $k$-vectors on the zone
edges either, the group theory dictates that the eigenstates on those
$k$-points should have the parity 1 or -1.  Then we can always
construct the wave function for one half of the unit cell by multiplying 
the wave function for the 
other with 1 or -1.  Namely, if we employ 
a simple cubic unit cell for the  P-surface (Fig.*, 
which is twice the bcc unit cell 
depicted in Fig.\ref{fig:SurfaceUnit}(a)) to make the 
correspondence clearer, an application of $\sigma_p$
to the upper half?, the lower turns out to have just the same connection
number s? as in the D-surface as shown? in Fig.\ref{fig:Diagram}(b).  
So we have now the same connection numbers between P and D, where  
the only difference is different connection coefficients 
($\rho_i$'s) between P and D as indicated in the figure.  
If we compare these, we end up with a `law of corresponding 
$k$-points', 

\begin{tabular}{clcc}
P-surface & & & D-surface\\ 
$k$-points$(\rho_x,\rho_y,\rho_z)$ & & & 
$k$-points$(\rho_x,\rho_y,\rho_z)$ \\ \hline

$\Gamma$(1,1,1) & $\sigma_p = 1$ & $\Leftrightarrow$ & $\Gamma$(1,1,1)\\
 & $\sigma_p = -1$ & $\Leftrightarrow$ & R(-1,-1,-1)\\ \hline
M(-1,-1,1) & $\sigma_p = -1$ & $\Leftrightarrow$ & X(1,1,-1)\\
 & $\sigma_p = 1$ & $\Leftrightarrow$ & M(-1,-1,1)
\end{tabular}

We can immediately translate this into the correspondence 
found above, 
$\Gamma$, H $\Leftrightarrow$  $\Gamma$, R, etc, 
if we note the relation, 

\begin{tabular}{ccc}
simple cubic & & bcc\\ \hline
$\Gamma$ & $\Leftrightarrow$ & $\Gamma$, H\\
M & $\Leftrightarrow$ & N
\end{tabular}\\
between the simple-cubic and bcc unit cells.

The wave functions shown in Figs.\ref{fig:WaveP_D}, 
are actually related through this relation.  
A simplest way to confirm this is to note that the wave function 
on each unit patch behaves in a similar manner.  In fact,  
the wave function $\psi(\theta, \phi)$ 
in eqn.\ref{eqn:SchOnP} is identical between the 
two surfaces. 

\section{Discussions}
The band structures revealed here should have
important implications on various physical properties.  
These should include transport properties as well as 
the cyclotron resonance, which can detect the effective mass
arising from topological band structures.  
Since the mass is determined by the interference of 
wave functions, effects of external magnetic fields 
should also be interesting. 

We can finally comment that, if we adopt foams of
graphite to realize curved surfaces\cite{MacTerr},
then the equation of motion of $\pi$ electrons on the network of
the honeycomb lattice will become, in the effective mass picture,
the problem of zero-mass Dirac equation
(i.e., Weyl's equation) on curved surfaces.
While we have ignored spin degrees of freedom here,
the spin connection on the surface will give rise to
a Berry's geometrical phase.


H.A. wishes to thank 
Alan Mackay for illuminating correspondences, 
Hiroshi Kuratsuji and Koichi Fukuda for discussions, 
and Yasuo Nozue for pointing out the ref.\cite{kyotani}.

\begin{figure}
\caption{Two ways to prepare a surface: (a) to introduce potential barriers
that confine electrons in a three-dimensional space to a thin membrane,
or (b) to roll a sheet of free electrons into the curved surface. }
\label{fig:Schematic}
\end{figure}

\begin{figure}
\caption{
The structures of P(a) and D(b) surfaces.  
We show a unit patch on the left panel, and a full unit cell 
on the right.  The grey-scale in the right panel 
represents the curvature potential, where 
shaded (open) circles depict 
potential minima (maxima, coinciding with the navels).
(c) The stereographic projection from minimal surface to Gauss spheres. 
A unit patch of P or D corresponds to a pair of 1/8 spheres.
}
\label{fig:SurfaceUnit}
\end{figure}

\begin{figure}
\caption{
Descretization for the spherical coordinates: 
(a) a mesh with even intervals in $\theta$ and $\phi$
and (b) uneven ones, adopted here, which take 
care of the singular point in the Jacobian.
}
\label{fig:pointP}
\end{figure}

\begin{figure}
\caption{
  The energy band structure
  in units of $\hbar^2/2mL^2$ is shown for the P-surface,
  when the curvature potential is considered (curves)
  or ignored (dots) with an energy offset to make the band bottoms
  coincide between the two cases. The inset depicts the Brillouin 
  zone for a bcc unit cell.
  }
\label{fig:BandP}
\end{figure}

\begin{figure}
\caption{
Typical wavefunctions for a unit cell of P-surfaces 
with positive (negative) amplitudes color-coded in red
(blue). Their eigenenergies are indicated in Fig.\ref{fig:BandP}.
}
\label{fig:WaveP}
\end{figure}

\begin{figure}
\caption{
  The band structures for the P- and D-surfaces.
  Horizontal lines indicate how the energies at the zone center or
  edges coincide between the two cases.
  }
\label{fig:BandP_D}
\end{figure}

\begin{figure}
\caption{
Typical wavefunctions for a unit cell of the Bonnet-connected 
P- and D-surfaces with positive (negative) amplitudes color-coded in red
(blue). Their eigenenergies are indicated in Fig.\ref{fig:BandP_D}.
}
\label{fig:WaveP_D}
\end{figure}

\begin{figure}
\caption{
(a)The way in which patches (labeled by large 
numbers) are connected to those in 
adjacent unit cells are indicated by 
small numbers attached to the edges for 
the D-surface. The Bloch
phase factors ($\rho_i$'s) are also shown. 
In the left panels the patches are flattened and expanded, 
while the right panels depict the actual three-dimensional shapes.
(b)The corresponding diagram for the P-surface,
where we have rearranged the numbers to 
make them identical with those for the D-surface 
by exploiting the symmetry. Accordingly the Bloch
phase factors for the P-surface involve the parity 
inversion ($\sigma_p$). 
An example is shown in (c), which indicates 
how patches 7,8 are neighboring 
3,4 through $\rho_x \sigma_p$.
}
\label{fig:Diagram}
\end{figure}

\end{multicols}

\begin{thebibliography}{99}

\bibitem[*]{Takeda} Present address: 
SGI Japan, Ltd., Ebisu, Shibuya, Tokyo 150-6031, Japan

  \bibitem{dewitt} B.S. DeWitt,
   Rev. Mod. Phys. {\bf 29}, 377 (1957).

  \bibitem{DiffG} See, e.g., S. Kobayashi and K. Nomizu:
  {\it Foundations of Differential Geometry} (John Wiley, 1969).

  \bibitem{mackayperi} A.L. Mackay,
  Nature {\bf 314}, 604 (1985).

  \bibitem{Mackay1} H. Terrones and A.L. Mackay
  in {\it Growth Patterns in Physical Sciences and Biology} ed. by
  J.M. Garcia-Ruiz et al (Plenum, New York, 1993) p.315.

  \bibitem{Mackay2} A. Mackay,
Current Science {\bf 69}, 151 (1995).

  \bibitem{enume} O.D. Friedrichs et al.,
  Nature {\bf 400}, 644 (1999).

  \bibitem{MacTerr} A.L. Mackay and H. Terrones, Nature {\bf 352}, 762
(1991).

  \bibitem{melano} S. Tsuneyuki et al.
study.
  in M. Doyama et al (eds): {\it Computer Aided Innovation of
  New Materials} (Elsevier, 1991), p. 381.

  \bibitem{PMM} 
H. Aoki, Y. Syono and R. J. Hemley (eds): 
{\it Physics Meets Mineralogy --- Condensed-Matter Physics 
in Geosciences}, (Cambridge Univ. Press, 2000).

  \bibitem{moriguchi} K.Moriguchi et al., Phys.Rev. B {\bf 61}, 9859
(2000)
  and refs therein.

  \bibitem{Tersoff}
  D. Vanderbilt and H. Tersoff, Phys. Rev. Lett. {\bf 68}, 511 (1992).

  \bibitem{OKeef}
  M. O'Keeffe, Phys. Rev. Lett. {\bf 68}, 2325 (1992).

  \bibitem{Lenosky}
  T. Lenosky et al., Nature {\bf 355}, 333 (1992).

  \bibitem{Fujita}
  This kind of structures has been generalized into what M. Fujita
et al [Phys. Rev. B {\bf 51}, 13778 (1995)]
call `Pearcene' after the architect Peter Pearce.

  \bibitem{kyotani} T. Kyotani et al., Chem. Mater. {\bf 1997}, 609
(1997).

  \bibitem{Schwarz} H.A. Schwarz, {\it Gesammelte Mathematische
  Abhandlungen} (Springer, Berlin, 1890).

  \bibitem{Nagaoka}
   M. Ikegami and Y. Nagaoka,
Prog. Theoret. Phys. Suppl. No.106, 235 (1991).

  \bibitem{Ogawa} N. Ogawa et al.,
Prog. Theoret. Phys. {\bf 83}, 894 (1990).

  \bibitem{Takagi}
    M. Ikegami et al.,
{\it Prog. Theoret. Phys.} {\bf 88}, 229 (1992).

  \bibitem{Terrones}
    H. Terrones,
    J. de Physique, Colloque C7, {\bf 51}, 345 (1990).

  \bibitem{Cvijovic}
    D. Cvijovi\'{c} and J. Klinowski,
    J. de Physique {\bf 2}, 2191 (1992).

  \bibitem{Klinowski}
    J. Klinowski, A.L. Mackay, and H. Terrones,
     Phil. Trans. R. Soc. Lond. A {\bf 354}, 1975 (1996).

  \bibitem{Hilbert}
  D. Hilbert and S. Cohn-Vossen, {\it Anschauliche Geometrie}
  (Springer, Berlin, 1932), \S 28.

  \bibitem{foam}
  If we mimic such structures with an assembly of atoms, the potential
maxima
around the neck correspond to the positions of eight-membered rings
(introduced to make the curvature negative)
on the `graphite foam'.\protect\cite{Mackay1,Mackay2,MacTerr}

  \bibitem{Townsend} S.J. Townsend et al.,
  Phys. Rev. Lett. {\bf 69}, 921 (1992).

  \bibitem{foam}
If we mimic such structures with a graphite foam
(i.e., a membrane of mostly hexagonal lattice), the potential minima
correspond to the positions of eight-membered rings
(introduced to make the curvature negative)
\protect\cite{Mackay1,MacTerr,Townsend}.

  \bibitem{hyde} S.T. Hyde and S. Andersson,
  Z. Kristall.  {\bf 174}, 225 (1986).  However, we cannot actually 
  change the Bonnet angle continuously, since 
the surface does not close and intersects with itself 
for the angles between these values.  

\end{thebibliography}
\end{document}